\documentclass[aps,prl,preprint,superscriptaddress,showpacs,showkeys]{revtex4}
\usepackage{epsfig}
\usepackage{amsmath}
\newcommand{\smartpap}{p\hskip-7pt\hbox{$^{^{(\!-\!)}}$}}

\begin{document}
\preprint{FNT/T 2003/05}

\title{Higher-order QED corrections to $W$-boson
mass determination at hadron colliders}

\author{C.M.~Carloni Calame}
\affiliation{Istituto Nazionale di Fisica Nucleare, Sezione di Pavia, 
via A. Bassi 6, I-27100, Pavia, Italy}
\affiliation{Dipartimento di Fisica Nucleare e Teorica, 
Universit\`a di Pavia, via A. Bassi 6, I-27100, Pavia, Italy}
\author{G. Montagna}
\affiliation{Dipartimento di Fisica Nucleare e Teorica, 
Universit\`a di Pavia, via A. Bassi 6, I-27100, Pavia, Italy}
\affiliation{Istituto Nazionale di Fisica Nucleare, Sezione di Pavia, 
via A. Bassi 6, I-27100, Pavia, Italy}
\author{O. Nicrosini}
\affiliation{Istituto Nazionale di Fisica Nucleare, Sezione di Pavia, 
via A. Bassi 6, I-27100, Pavia, Italy}
\affiliation{Dipartimento di Fisica Nucleare e Teorica, 
Universit\`a di Pavia, via A. Bassi 6, I-27100, Pavia, Italy}
\author{M. Treccani}
\affiliation{Dipartimento di Fisica Nucleare e Teorica, 
Universit\`a di Pavia, via A. Bassi 6, I-27100, Pavia, Italy}

\date{\today}

\begin{abstract}
The impact of higher-order final-state photonic corrections on the precise 
determination of the $W$-boson mass at the Tevatron and LHC
colliders is evaluated. In the presence 
of realistic selection criteria, the shift in the $W$ mass 
from a fit to the transverse mass distribution is found 
to be about 10~MeV in the $W \to \mu \nu$ channel and 
almost negligible in the $W \to e \nu$ channel. 
The calculation, which is implemented in a
Monte Carlo event generator for data analysis, 
can contribute to reduce the uncertainty
associated to the $W$ mass measurement at future 
hadron collider experiments. 
\end{abstract}


\maketitle

Precision tests of the Standard Model 
require a more and more accurate knowledge of the
basic parameters of the theory. In particular, future measurements
of the $W$-boson and top quark masses at the Tevatron and the LHC colliders
are expected to considerably improve the present indirect 
bound on the Higgs-boson mass from electroweak 
precision data. As recently discussed in the literature~\cite{snow2001}, 
a precision of 27~MeV (16 MeV) for the $W$ mass $M_W$ is the target
value for Run IIa (Run IIb) of the Tevatron. An accuracy of 
15 MeV is the final goal of LHC~\cite{lhc}.

In order to measure $M_W$ with such a high precision in a hadron
collider environment, it is mandatory to keep under control higher-order 
QCD and electroweak radiative corrections to the $W$ and $Z$
production processes. The status of QCD corrections to weak
boson production in hadronic collisions is reviewed in 
Ref.~\cite{qcddy}, while recent progress in the calculation of
electroweak corrections, as achieved by means of
independent calculations~\cite{ewo,bkw,ew-nc}, is 
summarized in Ref.~\cite{ewdy}. As shown in Refs.~\cite{ewo,bkw,ew-nc,ewdy}, 
electroweak corrections are dominated by photon radiation
effects and, in particular, by final-state photon emission, which
gives rise to collinear logarithms of the form 
$\alpha/\pi \log (\hat{s}/m_l^2)$,
where $\hat{s}$ is the effective centre of mass (c.m.) energy 
and $m_l$ is the mass of the final-state lepton. 
This poses the question of the impact of 
higher-order ({\it i.e.} beyond order $\alpha$) leading logarithmic 
corrections due to multi-photon radiation. A first attempt 
toward the inclusion of ${\cal O}(\alpha^2)$ QED corrections 
was the calculation of the double-bremsstrahlung 
matrix elements $q \bar{q'} \to W \to l \nu \gamma \gamma$ 
and $q \bar{q} \to \gamma,Z \to l^+ l^- \gamma \gamma$ 
($l=e,\mu$) performed in Ref.~\cite{bs}.
The aim of the present work is to evaluate the impact of higher-order
final-state QED corrections on the $W$ mass determination 
at hadron colliders, by including 
both real bremsstrahlung and virtual corrections. To this end, 
a Parton Shower (PS) approach in QED~\cite{ps-pv} is employed to simulate 
multi-photon radiation effects. An independent calculation 
of multi-photon radiative corrections in leptonic $W$ decays
has appeared very recently~\cite{pj}, but without quantifying their impact 
on the $W$ mass measurement. The uncertainty in the $W$ mass due to
higher-order QED effects is presently estimated by the CDF 
collaboration at the Tevatron to be 20 MeV in the $W \to e \nu$ channel, 
and 10 MeV in the $W \to \mu \nu$ channel~\cite{cdf}. 
An uncertainty of 12 MeV is assigned by the D\O\ collaboration
to the $W \to e \nu$ channel~\cite{d0}. 

From our analysis, we find that the shift in the fitted $W$ mass
is about 10~MeV in the $W \to \mu \nu$ channel and almost negligible
in the $W \to e \nu$ channel. The inclusion of the present
calculation in the experimental analysis would reduce significantly 
such a theoretical uncertainty in 
future improved measurements of the $W$ mass at hadron colliders.

An appropriate theoretical tool to compute photonic
radiative corrections in the leading log approximation is the
QED PS approach~\cite{ps-pv}. It consists in a numerical solution
of the QED Gribov-Lipatov-Altarelli-Parisi evolution equation 
for the charged lepton Structure Function $D(x,Q^2)$ in the non-singlet channel.
The solution can be cast in the form~\cite{ps-pv}
\begin{eqnarray} 
&& D(x,Q^2)=\Pi(Q^2,m^2)\delta(1-x)\nonumber\\
&+&\bigg(\frac{\alpha}{2\pi}\bigg) \int_{m^2}^{Q^2} \! \Pi (s,s')\frac{d s'}{s'}
\Pi (s',m^2)
\int_0^{x_+} \! dy P(y) \delta (x-y)\nonumber\\
&+&\bigg(\frac{\alpha}{2\pi}\bigg)^2\int_{m^2}^{Q^2}\Pi (s,s')
\frac{ds'}{s'}\int_{m^2}^{s'}\Pi (s',s'')\frac{ds''}{s''}
\Pi (s'',m^2) \nonumber \\
&&\int_0^{x_+} dx_1\int_0^{x_+}dx_2 P(x_1)P(x_2) \delta (x-x_1x_2) + \cdots    
\label{eq:alpha2}             
\end{eqnarray}
where $\Pi (s_1,s_2) = \exp \left[-\frac{\alpha}{2 \pi} 
\int_{s_2}^{s_1} \frac{d s'}{s'} \int_0^{x_+} dz P(z)  \right]$ 
is the Sudakov form factor, $P(z)$ is the $e \to e + \gamma$ 
splitting function and $x_+$ is an infrared regulator, 
separating the soft+virtual region from hard bremsstrahlung.
Equation (\ref{eq:alpha2}) allows to compute $D(x,Q^2)$
by means of a Monte Carlo algorithm which, as shown in Ref.~\cite{ps-pv}, 
simulates the emission of a shower of (real and virtual) photons by a
charged fermion and accounts for
exponentiation of soft photons and re-summation of collinear logarithms due
to multiple hard bremsstrahlung. A remarkable advantage of the PS algorithm 
with respect to a strictly collinear approximation is the possibility 
of generating transverse momentum $p_T$ of fermions and photons at each
branching, thus allowing an exclusive event generation suitable to
implement experimental cuts according to a realistic event selection.
The generation of transverse degrees of freedom can be performed according
to different recipes, as described in detail in Ref.~\cite{ps-pv}. 
Here, we generate photon angular variables according to the
leading pole behavior $1/(1 - \beta_l \cos\vartheta_{l\gamma})$,
where $\beta_l$ is the lepton velocity and $\vartheta_{l\gamma}$
is the relative lepton-photon angle.

A simple recipe to evaluate final-state corrections to
$p\smartpap \to W \to \nu l$ consists in attaching a single 
structure function $D(x,Q^2)$ to the lepton coming from the $W$ decay. Needless to
say,  this amounts to neglect photonic corrections due to initial-state
radiation, initial-final-state interference and $W$-boson emission. However, it
is known that radiation
from an internal off-shell particle can not contribute
to leading logarithmic corrections, which are the
main concern of the present study. Further, initial-state
photon radiation requires an appropriate
treatment, since radiation off quarks gives rise to quark
mass singularities which, as discussed in Refs.~\cite{ewo,bkw,ew-nc,ewdy}, can
be reabsorbed in Parton Distribution Functions (PDF), 
in analogy to gluon emission in QCD. These considerations imply that the
treatment of final-state photon radiation alone is not gauge invariant.
Nevertheless, it can be easily checked, by comparing the PS spectrum
with the gauge-invariant factor for collinear photon emission 
by a fermion~\cite{coll}, that gauge violations
are confined to the next-to-leading order corrections, which are
beyond the approximation of the present analysis. This issue is
further discussed in the following, by means of a quantitative
comparison with the independent, gauge-invariant calculation
of Ref.~\cite{bkw}.

\begin{table}
\caption{\label{comparison} Comparison between the 
present calculation (HORACE) and WGRAD~\cite{bkw,wgrad} 
for the $p \smartpap \to W \to l \nu$, $l=e,\mu$ cross 
sections (in pb), at the Tevatron Run II ($\sqrt{s}$ = 2 TeV) 
and the LHC ($\sqrt{s}$ = 14 TeV).}
\begin{ruledtabular}
  \begin{tabular}{l c c c c} 
     & \multicolumn{2}{c}{$\sqrt{s}$ = 2 TeV} 
     & \multicolumn{2}{c}{$\sqrt{s}$ = 14 TeV}\\ \cline{2-5}
     & $e$ & $\mu$ & $e$ & $\mu$\\
    \hline
    WGRAD Born
    & \multicolumn{2}{c}{441.7(1)} 
    & \multicolumn{2}{c}{1906(1)}\\    
    WGRAD                                   & 418.3(4) & 429.4(3) &
    1800(2) & 1845(2)\\
    WGRAD final-state                       & 419.7(1) & 430.0(1) & 
    1808(1) & 1854(1)\\
    HORACE Born
    & \multicolumn{2}{c}{441.6(1)} 
    & \multicolumn{2}{c}{1905(1)}\\ 
    HORACE ${\cal O}(\alpha)$               & 419.4(1) & 429.9(1) & 
    1806(1) & 1853(1)\\
    HORACE exponentiated                    & 419.5(1) & 430.0(1)& 
    1808(1) & 1853(1)\\
  \end{tabular}
  \end{ruledtabular}
\end{table}

In order to quantify the effect of higher-order final-state QED corrections
on the $W$ mass determination, we developed a Monte Carlo
event generator following the approach described above and performed a number of Monte Carlo 
experiments. The technical details of the Monte Carlo code HORACE
(Higher Order RAdiative CorrEctions) will be presented elsewhere. 
 Before the phenomenological analysis, we performed a tuned 
comparison between the predictions of HORACE and those of WGRAD~\cite{bkw,wgrad},
to verify the accuracy of our calculation. The results of such a comparison
are shown in Tab.~\ref{comparison}, using default PDFs, input parameters
and cuts as in Ref.~\cite{wgrad}. WGRAD includes the full set of 
${\cal O}(\alpha)$ electroweak radiative corrections to $W$ production 
(second line in
Tab.~\ref{comparison}) but it also gives the possibility to select
the effect of a gauge-invariant subset due to final-state corrections 
(WGRAD final-state in Tab.~\ref{comparison}).
Therefore, the difference, at a few per mille
level, between WGRAD and WGRAD final-state points out, 
when comparing with the Born predictions, the dominance of final-state radiation within
the full set of ${\cal O}(\alpha)$ electroweak corrections. On the other hand, 
it can be seen that the predictions by WGRAD final-state are in very good
agreement with our results by HORACE ${\cal O}(\alpha)$, which is an order $\alpha$
expansion of the complete PS algorithm. Since the differences are well below
the 0.1\% level, this comparison demonstrates that the gauge-invariance violations
present in our approach are numerically negligible. The contribution of
higher-order effects can be seen by comparing HORACE ${\cal O}(\alpha)$ with our
complete predictions given by HORACE exponentiated. Their effect on the
integrated cross section is tiny, within 0.1\%. For the sake of completeness, we 
performed also comparisons between WGRAD and HORACE at the level of differential
distributions, such as lepton transverse momentum and transverse mass
distributions, finding perfect agreement.

Having established the physical and technical accuracy of
our calculation, we move to the analysis of the $W$ mass 
shift due to higher-order corrections. The input parameters used in the 
simulations are: 
\begin{eqnarray}
&&m_{\nu_l} = 0 \qquad \quad m_e = 0.511 \times 10^{-3}~{\rm GeV} 
\nonumber \\
&& m_{\mu} = 0.10565836~{\rm GeV} \nonumber\\
&&\alpha^{-1} = 137.03599976  \quad G_{\mu} = 1.16639 
\times 10^{-5}~{\rm GeV}^{-2} \nonumber\\
&&\alpha_s = 0.1185 \nonumber \\
&&M_W = 80.423~{\rm GeV} \quad M_Z = 91.1882~{\rm GeV} \nonumber \\
&&\sin^2 {\theta_W} = 1 - \frac{M_W^2}{M_Z^2} \quad  
\Gamma_W = \frac{3 G_{\mu} M_W^3}{2\sqrt{2} \pi} 
\left(1 + \frac{2 \alpha_s}{3 \pi} \right) 
\end{eqnarray}
We adopt the $G_\mu$ scheme and fixed-width scheme in our calculation.
At the parton level, we consider the processes
\begin{eqnarray}
u + \bar{d} \to W^+ \to l^+ + \nu_l \quad
u + \bar{s} \to W^+ \to l^+ + \nu_l \nonumber\\
c + \bar{d} \to W^+ \to l^+ + \nu_l \quad 
c + \bar{s} \to W^+ \to l^+ + \nu_l
\end{eqnarray}
and their charge conjugate, with $l = e, \mu$ and CKM matrix elements 
according to Ref.~\cite{pdg}. The results for the 
processes $p \bar{p} \to W \to l + \nu$ (Tevatron)
and $p p \to W \to l + \nu$ (LHC) are obtained 
by convoluting the parton-level matrix element with 
CTEQ6 PDFs~\cite{cteq}. The virtuality scale $Q^2$ is set to be
$Q^2 = \hat{s}$, $\hat{s}$ being the effective c.m. energy
after gluon radiation, in both PDFs and lepton Structure Function.
The c.m. energies considered are $\sqrt{s} = 2$~TeV for the Tevatron and 
$\sqrt{s} = 14$~TeV for the LHC.

To model the acceptance cuts used by the CDF and D\O\ collaborations
in their $W$ mass analyses, we impose the following transverse momentum ($p_T$)
and pseudo-rapidity ($\eta$) cuts: 
\begin{eqnarray}
p_T(l) > 25~{\rm GeV}  \qquad |\eta(l)| < 1.2 
\qquad {\not\! p_T} > 25~{\rm GeV} 
\label{eq:cuts}
\end{eqnarray}
However, in order to perform a realistic phenomenological
analysis and study the dependence of the $W$ mass shift from detector
effects, we implement, in addition to
the above cuts, the lepton identification requirements quoted 
in Table I of Ref.~\cite{bkw}. According to these criteria, electron and 
photon four-momenta are recombined for small opening angles between
the two particles, consistently with a calorimetric
particle identification, while muons are identified as hits in the
muon chambers with an associated track consistent with a minimum 
ionizing particle. Furthermore, we simulate uncertainties in the
energy and momentum measurements of the charged leptons in the detector
by means of a Gaussian smearing of the particle four-momenta, using as 
standard deviation values the specifications 
relative to electrons and muons for the Run II D\O\ detector \cite{d0-det}.

\begin{figure}
\includegraphics[width=7.5cm]{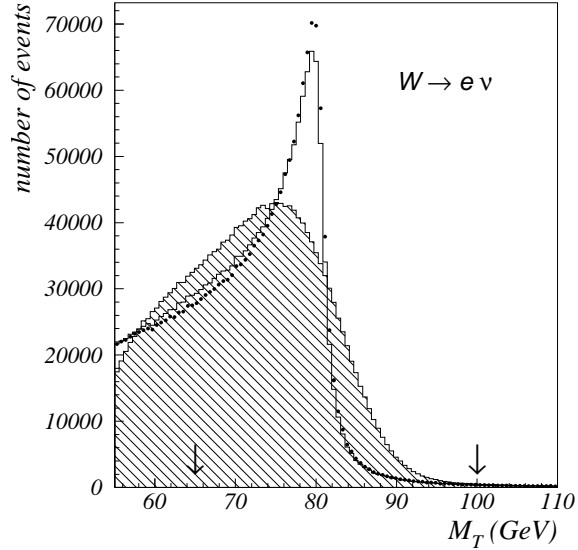}
\caption{\label{tmass} The transverse mass distribution 
without lepton identification criteria and detector 
resolutions (solid histogram), with lepton identification criteria 
(markers) and with detector resolutions (shaded histogram),
in the $W \to e \nu$ channel at $\sqrt{s} = 2$~TeV. Arrows indicate the
considered fit region.}
\end{figure}

The strategy followed by the CDF and D\O\ collaborations to extract $M_W$
from the data is to perform a maximum likelihood fit to the transverse mass distribution 
of the final-state lepton pair or to the transverse momentum of the charged
lepton. Here we consider the transverse mass, which is 
the preferred quantity to determine the $W$ mass and is defined as
\begin{equation}
M_T = \sqrt{2 p_T(l) p_T(\nu) (1 - \cos\phi^{l\nu})}
\end{equation}
where $p_T(l)$ and $p_T(\nu)$ are the transverse momentum of the 
lepton and neutrino, and $\phi^{l\nu}$ is the angle between the lepton 
and the neutrino in the transverse plane. The transverse mass distribution,
as obtained by our simulation, is shown in Fig.~\ref{tmass} at $\sqrt{s} = 2$~TeV.
The distribution without lepton identification requirements and
smearing effects (solid histogram) is compared to the distribution
including lepton identification criteria (markers) and detector
resolutions (shaded histogram). The shape of the $M_T$ spectrum is
considerably modified by detector resolution effects, 
in agreement with the results shown
in Refs.~\cite{cdf,d0}. The arrows in Fig.~\ref{tmass} select
the range $65~{\rm GeV} < M_T < 100$~GeV, which is used by CDF collaboration in its $W$
mass analysis and we also adopt in the fitting procedure described below.

\begin{figure}
\includegraphics[width=7.5cm]{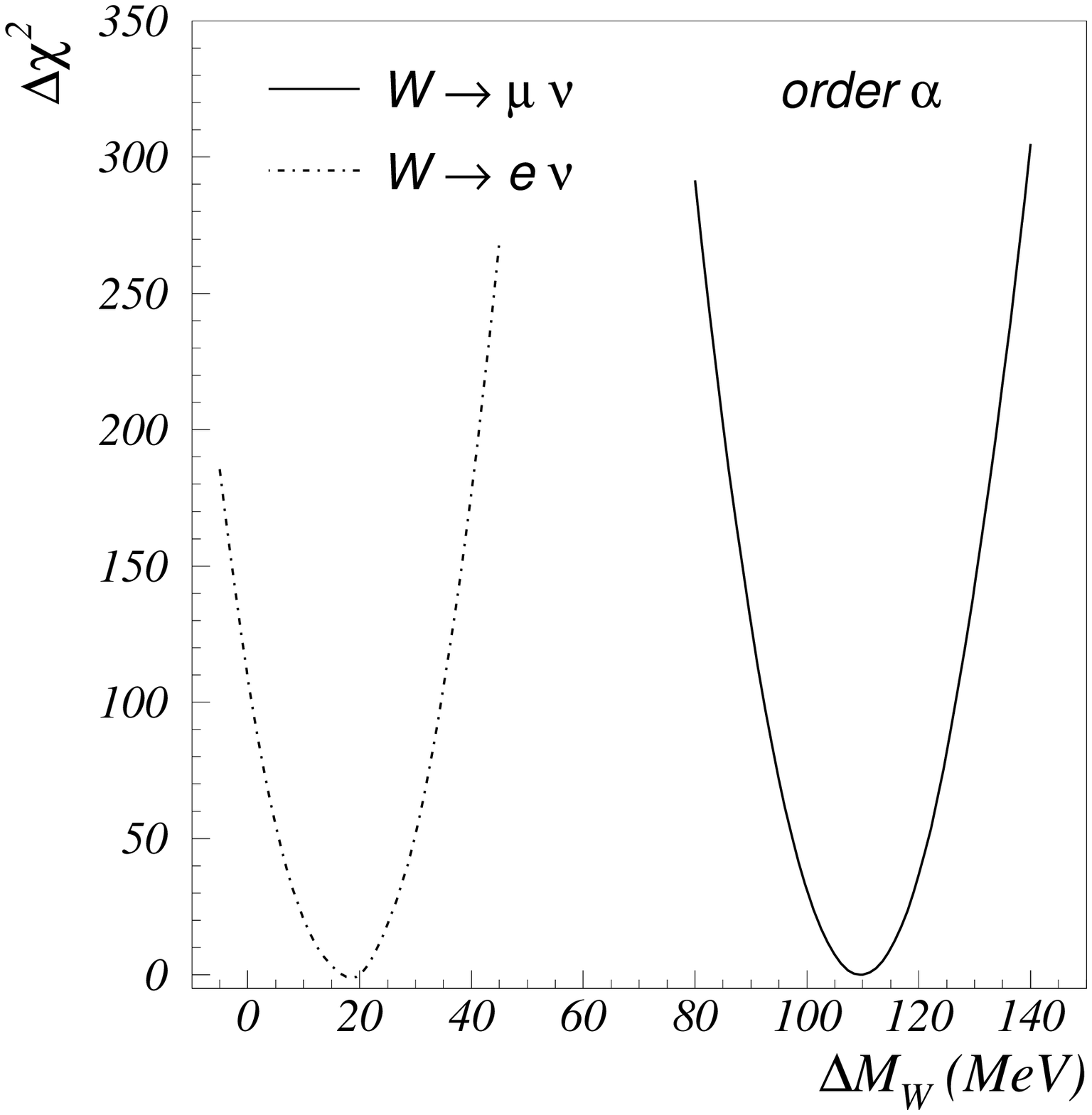}
\includegraphics[width=7.5cm]{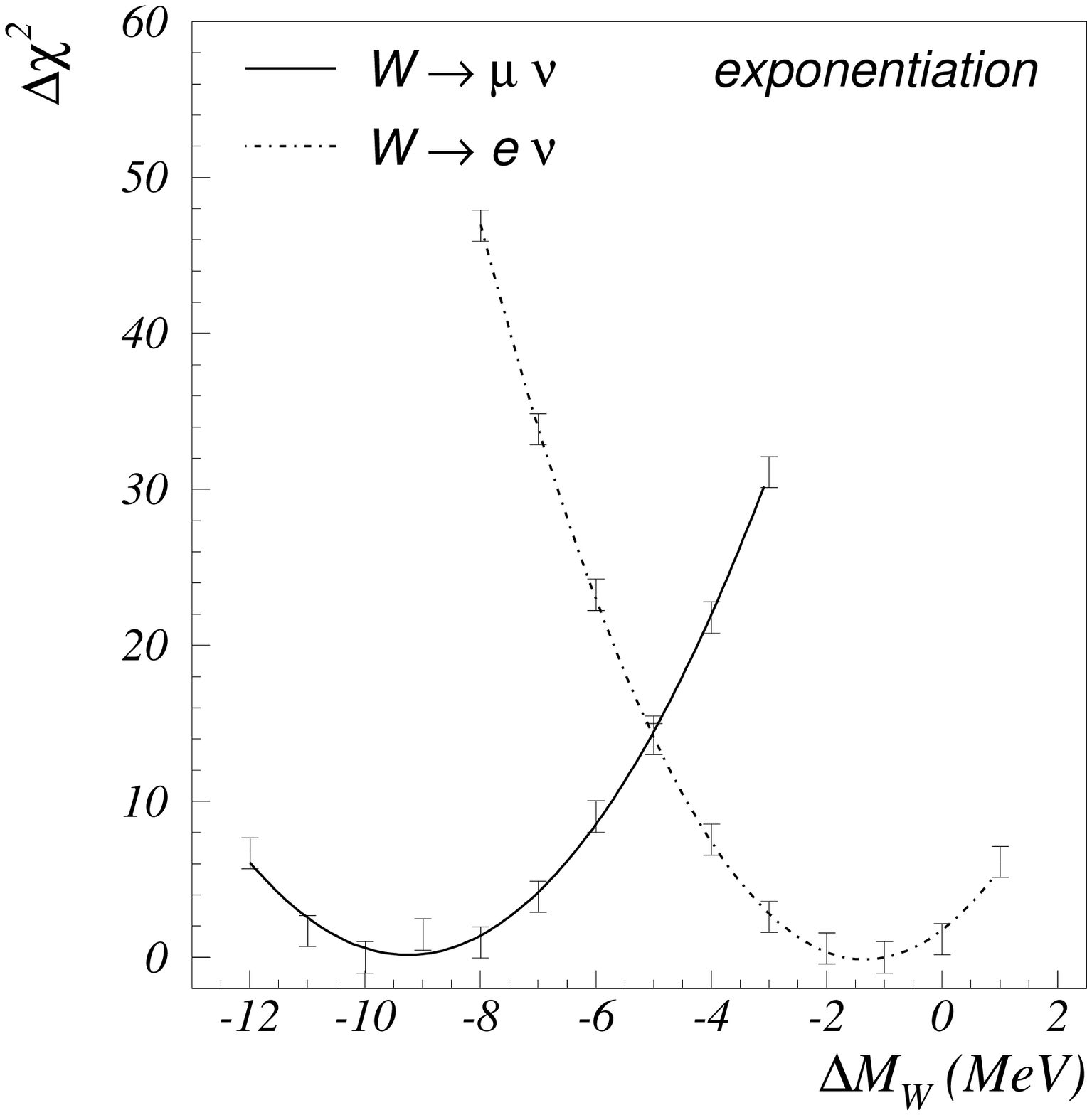}
\caption{\label{fit} The $\Delta\chi^2 = \chi^2 - \chi^2_{min}$ distributions 
from a fit to the $M_T$ distribution, including 
${\cal O}(\alpha)$ QED corrections
(left) and higher-order QED corrections (right), as 
a function of the $W$ mass shift, at $\sqrt{s} = 2$~TeV.
 The results for the 
 $W \to e \nu$ and $W \to \mu \nu$ channels are shown.}
\end{figure}

To evaluate the shift induced by higher-order corrections on the $W$
mass, we perform binned $\chi^2$ fits and binned 
maximum likelihood fits to the $M_T$ distribution, 
in complete analogy with the experimental fitting procedure.
Here we show only the results of the $\chi^2$ fits, because the results of the
maximum likelihood fits are in perfect agreement with the former. Using HORACE, we
generate a sample of pseudo-data and calculate with high numerical precision
the $m_T$ spectrum (binned into 100 bins) at the Born level and for a fixed,
"physical" value of the $W$ mass, {\it i.e.} $M_W^{ref}= 80.423$~GeV. Next, we
compute the $m_T$
spectrum including ${\cal O}(\alpha)$ 
leading-log corrections for 20 hypothesized $W$
mass values, with a spacing of 5 MeV for the $W \to  e \nu$ channel and
10 MeV for the $W \to \mu \nu$ channel. We then normalize the spectra within
the fit interval and we calculate, for each $M_W$ value, the $\chi^2$ as
\begin{equation}
\chi^2 \, = \, \sum_i (\sigma_{i,\alpha} - \sigma_{i,Born})^2/
(\Delta\sigma_{i,\alpha}^2 + \Delta\sigma_{i,Born}^2)
\end{equation}
where $\sigma_{i,Born}$ and $\sigma_{i,\alpha}$ are the Monte Carlo
predictions for the $i^{th}$ bin at the Born and ${\cal O}(\alpha)$ level,
respectively, and
$\Delta\sigma_{i,Born}, \Delta\sigma_{i,\alpha}$ the corresponding
statistical errors due to numerical integration. This allows to quantify
the mass shift due to ${\cal O}(\alpha)$ corrections. The shift due to
higher-order corrections is derived according to the same procedure, 
by generating a sample of pseudo-data for the $M_T$ distribution at
${\cal O}(\alpha)$ and fitting them in terms of the $M_T$ spectrum obtained including
higher-order corrections for 10 hypothesized $W$ mass values. In this case, we
use 1 MeV spacing between masses. Figure~\ref{fit} shows
the $\Delta\chi^2 = \chi^2 - \chi^2_{min}$ distributions as a function
of $\Delta M_W \equiv M_W - M_W^{ref}$, for the fit with 
${\cal O}(\alpha)$ corrections
(left) and the fit with higher-order corrections (right). The mass
shift observed for ${\cal O}(\alpha)$ corrections amounts to about $20$~MeV for the
$W \to e \nu$ decay (dashed line) and to $110$~MeV for the $W \to \mu \nu$
decay (solid line), as a consequence of the different
identification requirements. These shifts are 
in reasonable agreement with the results of the CDF and D\O\
collaborations, even in the absence of a complete detector simulation. The mass
shift due to higher-order effects is about $10$~MeV for the $W \to \mu \nu$ channel
(solid line) and a few MeV (dashed line) for the $W \to e \nu$ channel.
We performed the same analysis for the LHC collider (using the cuts
and pseudo-detector simulation of the Tevatron collider) and found   
that the same conclusions do apply to the LHC.

In conclusion, we have evaluated the impact 
of higher-order final-state QED corrections
on the determination of the $W$ mass at hadron
colliders, in view of future improved measurements
with an accuracy of 15-30 MeV. In the presence of realistic selection 
criteria, we have found that the shift 
due to these corrections is about 10~MeV in the
$W \to \mu \nu$ channel and practically negligible in the $W \to e \nu$ channel.
The calculation, if included in future experimental analyses, 
would reduce the uncertainty in the precision measurement of the 
$W$ mass at hadron colliders.
To this end, the Monte Carlo program HORACE is available 
for data analysis. 
A more realistic analysis would require a full detector simulation, which
is beyond the scope of the present paper.
The study of the neutral-current process 
$p\smartpap \to \gamma,Z \to l^+ l^-$ is left to a future work.


\end{document}